\documentclass[ag]{copernicus}  %82 pp a 1500

\usepackage[T1]{fontenc}
\usepackage{natbib}
\usepackage{color}

\begin{document}

\title{The usefulness of Poynting's theorem in magnetic turbulence 
}

\author[1]{Rudolf A. Treumann\thanks{Visiting scientist at the International Space Science Institute, Bern, Switzerland\\ \\ \emph{Correspondence to}: W. Baumjohann (Wolfgang.Baumjohann@oeaw.ac.at)}}
%\author[3]{R. Nakamura}
\author[2]{Wolfgang Baumjohann}
%\author[3]{James W. LaBelle}

\affil[1]{Department of Geophysics and Environmental Sciences, Munich University, Munich, Germany}
\affil[2]{Space Research Institute, Austrian Academy of Sciences, Graz, Austria}
%% The [] brackets identify the author to the corresponding affiliation, 1, 2, 3, etc. should be inserted.
%\affil[3]{Department of Physics and Astronomy, Dartmouth College, Hanover NH 03755, USA}

\runningtitle{Poynting's theorem in MHD}

\runningauthor{R. A. Treumann, W. Baumjohann}

\received{ }
%\pubdiscuss{ } %% only important for two-stage journals
\revised{ }
\accepted{ }
\published{ }

%% These dates will be inserted by the Publication Production Office during the typesetting process.

\firstpage{1}

\maketitle

\begin{abstract}
%\noindent\textbf{Abstract}. -- 
{We rewrite Poynting's theorem, already used in a previous publication \citep{treumann2017} to derive relations between the turbulent magnetic and electric power spectral densities, to make explicit where the mechanical contributions enter. We then make explicit use of the relativistic transformation of the turbulent electric fluctuations to obtain expressions which depend only on the magnetic and velocity fluctuations. Any electric fluctuations play just an intermediate role. Equations are constructed for the turbulent conductivity spectrum in Alfv\'enic and non-Alfv\'enic turbulence in extension of the results in the above citation. An observation-based discussion of their use in application to solar wind turbulence is given. The inertial range solar wind turbulence exhibits signs of chaos and self-organisation.}   
\end{abstract}

\vspace{0.5cm}
\section{Introduction}
{In a recent communication \citep{treumann2017} we used Poynting's theorem in electrodynamics in order to construct an experimentally accessible expression for the spectral energy density of the electromagnetic field in collisionless magnetic turbulence. That attempt turned out much simpler and therefore also more effective than our previous fairly involved inverse scattering theory \citep{treumann2016b} of electromagnetic fluctuations in magnetic turbulence.} Since we used only electromagnetic theory {not referring to any mechanical} fluid turbulence, it remained unclear to what extent an approach in turbulence like that one was justified. Magnetic turbulence at low frequencies -- scales longer than the electron gyro-radius -- involves \emph{both} the electromagnetic \emph{and} mechanical flow fields. {Restricting to one of these components only, apparently neglects an important part of the turbulence.} This argument also applies to any experiments which use just measurements of magnetic fluctuations, calculate spectral energy densities and possibly do not refer to electric field respectively velocity fluctuations. Determination of the power law shape of those spectra contains information about the turbulence, but its physical content remains inaccessible. Spectral slopes are sensitive to varying physical conditions \citep{treumann2015}. Small changes in the slope, which {within experimental errors are difficult to detect}, may indicate completely different physics. 

Observations of magnetic turbulence in the solar wind take advantage of their easy accessibility in order to determine spectral slopes of the turbulent magnetic energy densities \citep[cf., e.g.,][for early reviews]{goldstein1995,zhou2004} in the frequency domain. {They enable distinguishing between Kolmogorov's \citep{kolmogorov1941,kolmogorov1962} spectral ranges of energy injection, constant energy flux, and dissipation in frequency space \citep[cf., e.g.,][and references therein]{alexandrova2009,brown2015,horbury2012,sahraoui2009,sahraoui2012,sahraoui2013,wicks2012}. Sometimes they enable distinction of Kolmogorov and Kraichnan regimes. They also provide absolute values of the turbulent magnetic energy density. Applying the Taylor hypothesis, limited information about the corresponding spatial scales has been obtained and, in a few cases, spectra of the electric field \citep{bale2005,chen2011} and streaming velocity fluctuations \citep{podesta2006,podesta2007,podesta2010,roberts2010,podesta2011a,podesta2011b,safrankova2013,safrankova2016} have been added. Measurements of turbulent density fluctuations in the solar wind \citep{celnikier1983,chandran2009,chen2012,safrankova2013,safrankova2016} have also been published.} 

{In the present note, following our previous attempt, Poynting's theorem is briefly re-examined in order to relate it to the inclusion of the mechanical part of turbulence and to clarify the effect of the electric and velocity fluctuations.}  

\section{Poynting's theorem in magnetic turbulence}
{Measurement of the Poynting flux in order to infer the plasma wave energy flow in near-Earth space has a long history. One of the first attempts \citep{labelle1992} was to determine its direction and absolute value in plasmaspheric electromagnetic ion-cyclotron waves. More recently it was used to detect dispersive whistlers in Earth's bow shock \citep{sundqvist2012} which are expected to contribute to shock reformation in quasiperpendicular shocks \citep[cf., e.g.,][for a rather complete account]{balogh2013} and to the investigation of the energy flow in kinetic Alfv\'en waves near the plasma sheet boundary \citep{stawarz2017} as a source of the auroral energy flow which often is attributed to the inflow of kinetic Alfv\'en waves \citep[cf., e.g.,][]{chaston2003} causing particle acceleration and radio emission \citep{labelle2002}. These works deal with the Poynting flux in particular waves only.} 

In magnetic/magnetohydrodynamic turbulence (at non-relativistic speeds) the equation of energy conservation, which is the generalization of Poynting's theorem in electrodynamics to the inclusion of mechanical energy transport, is quite generally written \citep{landau1998} in the form
\begin{equation}
\frac{\partial}{\partial t}\bigg(\frac{1}{2}\rho v^2 +\rho\varepsilon+\frac{B^2}{2\mu_0}\bigg)= -\nabla\cdot\vec{q}
\end{equation}
The vector $\vec{q}$ is the energy flux density, $\rho$ is the plasma mass density, $\vec{v}$ velocity, and $\varepsilon=w-P/\rho$ the internal energy, with $w$ internal enthalpy, $P=\mathrm{tr}\,\vec{\mathbf{P}}\equiv \frac{1}{3}P_{ii}$ (scalar) pressure, and the relativistically{-}small electric field density has been suppressed in the time-derivative term on the left. In this form the energy law accounts for all the energy in the turbulence. The energy flux vector $\vec{q}$ contains {all} the dissipative processes, mechanical and electromagnetic, in particular all anomalous processes which contribute to dissipation. The former {(mechanical) terms} contain a mechanical dissipation tensor, with bulk and shear viscosity coefficients. The latter {(electromagnetic) terms} are inherent to a conductivity tensor $\sigma_{ij}$, which enters Ohm's law and which can always be written in its simplest form, such that the current is given by ${J}_i=\sigma_{ij}E_j$, where $\vec{E}=-\vec{v\times B}$ is the (relativistically correct) electric field. {For finite electrical resistance, the relation between the electric field and current $\vec{J}$ becomes $\vec{E}+\vec{v\times B}=\vec{\sigma}^{-1}\cdot\vec{J}$, an expression which is general in the sense that the various dissipative processes contributing to this generalized Ohm's law \citep[cf., e.g.][]{baumjohann1996,krall1973} are included into the definition of the conductivity tensor $\vec{\sigma}$ {which in all realistic cases, if made explicit, becomes an involved expression}. Any dissipation of electromagnetic energy is given by the product $-\vec{E\cdot J}$.} Neglecting collisional dissipation, as is usually done in ordinary MHD, one has for
\begin{equation}
\vec{q}=\rho\vec{v}\Big(\frac{1}{2}v^2+w\Big) +\frac{1}{\mu_0}\vec{B\times(v\times B)}
\end{equation}
Written in terms of the electromagnetic field, energy conservation takes the form
\begin{equation}
\frac{\partial}{\partial t}\frac{B^2}{2\mu_0}=-\frac{1}{\mu_0}\nabla\cdot\Big(\vec{E\times B}\Big) -\nabla\cdot\vec{q}_m-\frac{\partial}{\partial t}\mathcal{E}_m
\end{equation}
with 
\begin{equation}
\vec{q}_m=\rho\vec{v}\Big(\frac{1}{2}v^2+w\Big), \qquad \mathcal{E}_m=\frac{1}{2}\rho v^2+\rho\varepsilon 
\end{equation}
Any possibly occurring dissipation is solely due to turbulent mixing and in this sense is `anomalous'. This is Poynting's theorem completed with the two mechanical terms on the right. On the left is the time variation of the magnetic energy density. The first term on the right is the divergence of the electromagnetic energy flux vector, a familiar quantity. The other two terms, depending on their signs either pump energy into the magnetic field by mechanical motion, as in the case of a dynamo, or dissipate magnetic energy. 

Since any dissipation of \emph{magnetic} energy, either positive or negative, can always be written as the above product $-\vec{E\cdot J}$, Poynting's theorem for the electromagnetic field under ideal dissipationless conditions in magnetic/magnetohydrodynamic turbulence can be written
\begin{equation}\label{eq-5}
\frac{\partial}{\partial t}\frac{B^2}{2\mu_0}= -\vec{E\cdot J}-\frac{1}{\mu_0}\nabla\cdot\Big(\vec{E\times B}\Big)
\end{equation}
which is its familiar version in electrodynamics {and}
\begin{equation}\label{eq-6}
{\vec{E\cdot J}= \nabla\cdot\vec{q}_m+\frac{\partial}{\partial t}\mathcal{E}_m + \vec{E\cdot\sigma^{-1}_{\mathit{an}}\cdot E}}
\end{equation}
{A possibly present anomalous conductivity $\vec{\sigma}_\mathit{an}$ caused by kinetic processes on scales shorter than the ion or electron inertial lengths respectively gyroradii $<\lambda_{i,e},r_{\mathit{ci,ce}}$ would appear as the last term in (\ref{eq-6}) but is not explicitly considered in the following.} In collisionless and non-viscous turbulent plasmas {the latter form applies at scales exceeding the Debye length and is far away from any molecular scale to which dissipation of the turbulent mechanical energy is attributed. In contrast,} the turbulent electromagnetic energy is {\emph{ultimately}} dissipated {at least already} at electron scales {$<\lambda_e,r_\mathit{ce}$} by spontaneous reconnection {\citep{treumann2015}} in small-scale current filaments.\footnote{{At scales shorter than the electron gyroradius electrons demagnetise and do not anymore contribute to magnetic fluctuations, electron thermal pressure does not balance the Lorentz force which contracts the current, and collisionless reconnection is spontaneous and explosive, causing electron exhausts, strongly deformed electron distributions and electron beams. Dissipation here is kinetic and electrostatic provided by plasma waves (Langmuir, ion sound, Bernstein, electron holes). Except for a possible filamentary Weibel mode which causes further filamentation of the current and turbulence, no non-radiative magnetic fields are generated here. Hence, the magnetic turbulence spectrum should decay at those scales. High frequency and thus weak radiative fields can be produced in addition by the electron cyclotron maser instability inside the exhaust.}} {These are generated progressively by turbulent self-organization in the spectral energy flow \citep{treumann2017} towards the short scales. There dissipation is anomalous, mediated by plasma-kinetic processes.}

\section{Application to turbulent fluctuations}
Writing all quantities as sums of mean fields plus fluctuations $F=\bar{F}+\delta F$ with average $\overline{F}=\bar{F}$ and $\overline{\delta F}=0$ in Eq. (\ref{eq-5}), averaging, subtracting the scale averaged equation, and dropping the averaged products of the fluctuations as these depend only on the mean-field scale, we find
\begin{eqnarray}
\frac{\partial}{\partial t}\bigg(\frac{2\vec{\bar{B}}\cdot\delta\vec{B}}{2\mu_0}+\frac{(\delta B)^2}{2\mu_0}\bigg)&=&\nonumber\\[-1.5ex]
\\[-1.5ex]
-\delta\vec{E}\cdot\bar{\vec{J}}-\delta\vec{E}\cdot\delta\vec{J}&-&\frac{1}{\mu_0}\nabla\cdot\Big[\delta\vec{E}\times\bar{\vec{B}}+\delta\vec{E}\times\delta\vec{B}\Big]\nonumber
\end{eqnarray}
With mean electric field $\vec{\bar{E}}=0$ {(see next section below)}, and defining $\delta\vec{J}=\vec{\sigma}^T\cdot\delta\vec{E}$, where $\vec{\sigma}^T$ is an equivalent turbulent conductivity tensor chosen such that it relates the turbulent current to the turbulent electric field, the mean current vanishes. The result is
\begin{eqnarray}\label{eq-7}
\frac{\partial}{\partial t}\bigg(\frac{2\vec{\bar{B}}\cdot\delta\vec{B}}{2\mu_0}+\frac{(\delta B)^2}{2\mu_0}\bigg)&=&\nonumber\\[-1.5ex]
\\[-1.5ex]
-\delta\vec{E}\cdot\vec{\sigma}^T\cdot\delta\vec{E}&-&\frac{1}{\mu_0}\nabla\cdot\Big[\delta\vec{E}\times\bar{\vec{B}}+\delta\vec{E}\times\delta\vec{B}\Big]\nonumber
\end{eqnarray}
which is the basic equation used in \citet{treumann2017}. {Restricting to magnetically non-compressive turbulence $\delta\vec{B}\cdot\bar{\vec{B}}=0$ makes the first term on the left vanishing. The first term in the brackets on the right vanishes for $\delta\vec{E}\|\vec{\bar B}$, the case $\vec{k}\perp\bar{\vec{B}}$ of propagation of the turbulent fluctuations perpendicular to the mean field. For parallel propagation this term contains the excluded compressive magnetic component. We are thus left with the simplified Poynting equation}
\begin{equation}\label{eq-8}
\frac{\partial}{\partial t}\frac{(\delta B_\perp)^2}{2\mu_0}=-\delta\vec{E}\cdot\vec{\sigma}^T\cdot\delta\vec{E} -\frac{1}{\mu_0}\nabla\cdot\Big[\delta\vec{E}\times\delta\vec{B}_\perp\Big]
\end{equation}
{All dynamics of the turbulent mechanical flow is implicit to $\vec{\sigma}^T$, %(Note that we consider only magnetically non-compressive turbulence here with $\delta\vec{B}_\|=0$.)  
%The relation to the turbulent flow and its dynamics remains involved. It is contained in the above used identification
%\begin{equation}
%\vec{J\cdot E}= \nabla\cdot\vec{q}_m+\frac{\partial}{\partial t}\mathcal{E}_m
%\end{equation}
%For the turbulent fluctuations this reduces to
%\begin{equation}
%\delta\vec{E}\cdot\vec{\sigma}^T\cdot\delta\vec{E} =\nabla\cdot\delta\vec{q}_m+\frac{\partial}{\partial t}\delta\mathcal{E}_m
%\end{equation}
which (keeping an anomalous conductivity $\vec{\sigma}_{\!\!\!\mathit{an}}$) is formally defined as}
\begin{equation}\label{sigma}
{\vec{\sigma}^T= (\delta\vec{E})^{-1}\cdot\bigg[\vec{\sigma}_{\!\!\!\mathit{an}}+\nabla\cdot\delta \vec{q}_m+\frac{\partial}{\partial t}\delta\mathcal{E}_m\bigg]\cdot(\delta\vec{E})^{-1}}
\end{equation}
where $\delta\vec{q}_m, \delta\mathcal{E}_m$ are the fluctuations of $\vec{q}_m,\mathcal{E}_m$. {Once, by the means of measuring the electromagnetic turbulent fluctuation spectrum, the turbulent conductivity spectrum $\vec{\sigma}^T_{\omega\vec{k}}$ has been determined as function of fluctuation frequency $\omega$ and wavenumber $\vec{k}$, its transformation back into real space provides a relation to the turbulent mechanical quantities.} 

\section{Turbulent electric and velocity fields}
{A difficulty arises in dealing with the electric field.  Relativistic invariance requires its transformation into the rest frame of the flow $\vec{E}'=\vec{E}+\vec{v\times B}$. In an ideal turbulent medium the moving frame speed depends on {the fluctuation} scale, which  in general  makes it difficult (if not impossible) to define a common moving frame valid on all scales. Splitting into mean and fluctuating quantities yields the averaged field
\begin{equation}
\overline{\vec{E}'} =\bar{\vec{E}} +\bar{\vec{v}}\times\bar{\vec{B}} +\overline{\delta\vec{v}\times\delta\vec{B}}
\end{equation}
which {in the moving frame must vanish}. This gives the mean electric field $\bar{\vec{E}}=-\bar{\vec{v}}\times\bar{\vec{B}}-\overline{\delta\vec{v}\times\delta\vec{B}}$. Measurement of the velocity fluctuations $\delta\vec{v}$ in the scale range of interest is required in the averaged second term. The fluctuating primed electric field becomes
\begin{equation}
\delta\vec{E}'=\delta\vec{E}+\delta\vec{v}\times\bar{\vec{B}}+\delta\vec{v}\times\delta\vec{B}+\bar{\vec{v}}\times\delta\vec{B}-\overline{\delta\vec{v}\times\delta\vec{B}}
\end{equation}
The mean magnetic field $\bar{\vec{B}}$ and the last averaged term are constant on the fluctuation scale. In an infinitely extended medium without boundaries the last term can be dropped yielding
\begin{equation}\label{eq-14}
\delta\vec{E}=-\delta\vec{v}\times\bar{\vec{B}}-\delta\vec{v}\times\delta\vec{B}-\bar{\vec{v}}\times\delta\vec{B}
\end{equation}
which is to be used in Eq. (\ref{eq-8}). It requires knowledge of the velocity fluctuations on the same scales (and with same resolution) as the magnetic fluctuations.  The second term on the right measures the `alignment' of the magnetic and velocity fluctuations.} 

{In so-called purely Alfv\'enic turbulence $\delta\vec{v}\|\delta\vec{B}$ and the (normalized to the total energy) cross helicity is close to unity, resulting in a linear relation for the fluctuating electric field $\delta\vec{E}=-\delta\vec{v}\times\bar{\vec{B}}-\bar{\vec{v}}\times\delta\vec{B}$. The electric fluctuations are perpendicular to both $\delta\vec{B}, \delta\vec{v}$ in this case. The Poynting flux vector term in Eq. (\ref{eq-8}) assumes the form $-\bar{\vec{B}}\cdot\nabla\big(\delta\vec{v}\cdot\delta\vec{B}_\perp\big)/\mu_0$ which eliminates the electric fluctuations in favour of the velocity field and reduces Eq. (\ref{eq-8}) to
\begin{eqnarray}\label{eq-14a}
\frac{\partial}{\partial t}\frac{(\delta B_\perp)^2}{2\mu_0}&=&-{\sigma}^T_\perp\bar{B}^2(\delta\vec{v})^2 -\frac{1}{\mu_0}\bar{\vec{B}}\cdot\nabla\big(\delta\vec{v}\cdot\delta\vec{B}_\perp\big)\nonumber\\[-1.5ex]
&&\\[-1.5ex]
&-&\sigma_\perp^T\Big[\bar{{v}}^2\big(\delta\vec{B}_\perp\big)^2-\big(\bar{\vec{v}}\cdot\delta\vec{B}_\perp\big)^2\Big]\nonumber
\end{eqnarray}
where $\sigma_\perp^T$ is the turbulent conductivity parallel to $\delta\vec{E}$, i.e. perpendicular to both $\delta\vec{B}_\perp,\bar{\vec{B}}$. The last terms contain only the mean flow components $\bar{\vec{v}}^\perp$ perpendicular to $\delta\vec{B}_\perp$. The complications they introduce disappear when transforming to the easily determined mean flow $\bar{\vec{v}}=0$. The Poynting term vanishes when considering spatial dependencies perpendicular to the mean field. More generally, since in Alfv\'enic turbulence $\delta\vec{v}=\alpha\delta\vec{B}_\perp$ with $\alpha$ some angular dependent scalar factor (which can, in principle, be determined from the fluctuations), the argument of the Poynting vector can be expressed through $(\delta{B}_\perp)^2$. Except for any spatial dependence of $\alpha$, the magnetic and velocity fluctuation spectra should thus be comparable in Alfv\'enic turbulence for either parallel or perpendicular propagation. (One may note that for cross-helicity $\delta\vec{v}\cdot\delta\vec{B}_\perp/|\delta\vec{v}\cdot\delta\vec{B}_\perp|\approx\pm 1$ the second term on the right in (\ref{eq-14a}) disappears.)} 

{Fourier transforming in space and time in the infinitely extended domain, assuming stationary and homogeneous conditions and constant $\alpha$ yields 
\begin{equation}\label{eq-15a}
\sigma^T_{\perp\omega\vec{k}}=\frac{i\omega}{2\mu_0\bar{B}^2}\Big(1-2\alpha\frac{\vec{k}\cdot\bar{\vec{B}}}{\omega}\Big)
\frac{(\delta B_\perp)^2_{\omega\vec{k}}}{(\delta v_\perp)^2_{\omega\vec{k}}}
\end{equation}
This holds in Alfv\'enic turbulence. (The contribution of a finite mean speed may be retained any time when wanted.) For cross-helicity one, the expression in parentheses in the first term on the right reduces to unity.\footnote{There is, of course, no obvious reason for $\alpha$ to be constant. In general it will depend on space and time which is suggested by the radial variation of the solar wind spectra with increasing solar distance  \citep{roberts2010}. Locally, the assumption of constancy is well justified however, as is also confirmed by solar wind observations at 1AU of the constancy of the cross helicity \citep{podesta2010}.} The turbulent response of the plasma contained in the conductivity spectrum $\vec{\sigma}^T_{\perp\omega\vec{k}}$ is, under stationary and homogeneous conditions, given by the ratio of the spectral energy densities of the turbulent magnetic and velocity fields. [We note in passing that this expression can also be exploited for constructing  \citep{treumann2017} a low frequency ``turbulent dispersion relation''  $\mathcal{N}^2\equiv{k^2c^2/\omega^2}=i\sigma^T_{\perp\omega\vec{k}}/\omega\epsilon_0$ which is not the solution of a linear eigenmode problem but determines the nonlinear relation between the turbulent frequencies $\omega$ and wavenumbers $\vec{k}$.]}

{For non-Alfv\'enic turbulence $\delta\vec{v}\perp\delta\vec{B}$, i.e. $\delta\vec{v}\cdot\delta\vec{B}=0$ which means that the cross-helicity vanishes. It is convenient to distinguish velocity fluctuations parallel and perpendicular to the mean field. If $\delta\vec{v}\|\bar{\vec{B}}$ the turbulent electric, magnetic and velocity fluctuations form a mutually orthogonal system $\delta\vec{E}=-\delta\vec{v}\times\delta\vec{B}_\perp$. Hence Poynting's vector becomes $\delta\vec{E}\times\delta\vec{B}_\perp= \delta\vec{v}_\|\big(\delta B_\perp\big)^2$, giving from (\ref{eq-8}) 
\begin{equation}\label{eq-15}
\frac{\partial}{\partial t}\frac{(\delta B_\perp)^2}{2\mu_0}=-{\sigma}^T_{\perp} (\delta{B}_\perp)^2(\delta{v}_\|)^2 -\frac{1}{\mu_0}\nabla_\|\Big[\delta{v}_\|(\delta{B}_\perp)^2\Big]
\end{equation}
for non-compressive non-Alfv\'enic magnetic turbulence.
It is obvious that in this case the cross-helicity contributes through the (parallel) divergence of the Poynting flux.} {Unlike the Alfv\'enic case, the last term in the above expression generally cannot be reduced further. Moreover, the first term on the right is a triple product, which makes any further treatment difficult.}

{If the turbulence is independent on the parallel direction such that the parallel turbulent wave vectors $k_\|=0$ vanish, then the last equation simplifies and can be solved for the perpendicular non-Alfv\'enic conductivity spectrum:
\begin{equation}
\sigma_{\perp\omega{\vec{k}_\perp}}^T=\frac{i\omega}{2\mu_0}\Big[\log\big(\delta B_\perp\big)^2\Big]_{\omega{\vec{k}_\perp}}\Big[\big(\delta v_\|\big)^2_{\omega{\vec{k}_\perp}}\Big]^{-1}
\end{equation}
The logarithmic dependence on the spectral energy density of the magnetic turbulence implies that the conductivity spectrum is mainly determined by the spectral energy density in the turbulence of the  mechanical flow. This is also the case when $k_\|\neq0$, because then the above equation can be brought into the form 
\begin{equation}
\Big(\frac{\partial}{\partial t}+2\delta v_\|\nabla_\|\Big)\frac{\log\big(\delta B_\perp\big)^2}{2\mu_0}=-\frac{1}{\mu_0}\nabla_\|\delta v_\| -\sigma_\perp^T\big(\delta v_\|\big)^2
\end{equation}
where the dependence on the magnetic fluctuation spectrum remains to be logarithmic as well. Again, in homogeneous stationary turbulence this can be reduced to an equation for the spectral density of $\sigma_\perp^T$.} 

{Otherwise, for $\delta\vec{v}\perp\bar{\vec{B}}$, one has $\delta\vec{B}\|\bar{\vec{B}}$ as consequence of $\delta\vec{v}\perp\delta\vec{B}$. We called this case compressive magnetic turbulence \citep{treumann2017} and, for our purposes, excluded it from consideration.}

{Further conclusions can be drawn when considering the propagation of the turbulent fluctuations. Propagation perpendicular to $\bar{\vec{B}}$ of magnetically non-compressive fluctuations ($\delta\vec{B}_\|=0$) implies $\delta\vec{E}\|\bar{\vec{B}}$. Hence the first term on the right in Eq. (\ref{eq-14}) is zero, and since the magnetic and electric fluctuation fields are orthogonal, lying both in the plane perpendicular to the mean field, one has $\delta\vec{v}\|\bar{\vec{B}}$, i.e. all velocity fluctuations which contribute are parallel to the mean field. Moreover, in Eq. (\ref{eq-15}) the last term on the right thus disappears and, after Fourier transformation, one obtains a simple expression for the turbulent conductivity spectrum in homogeneous stationary turbulence in this case \citep{treumann2017}.}

{Any magnetically compressive turbulence $\delta\vec{B}\|\bar{\vec{B}}$, which so far has been excluded here, requires a separate investigation. In this case, still considering only electromagnetic fluctuations with $\delta\vec{E}\cdot\delta\vec{B}=0$ the electric fluctuations corresponding to $\delta\vec{B}_\|$ are perpendicular to $\bar{\vec{B}}$ in agreement with (\ref{eq-14}). One obtains after some simple algebra that
 \begin{equation}
\delta\vec{E}\times\delta\vec{B}_\|=\delta b_\|(1+\delta b_\|)\bar{B}^2\delta\vec{v}_\perp^\perp
\end{equation}
where $\delta\vec{v}_\perp^\perp$ is the velocity fluctuation perpendicular to the mean magnetic and the turbulent electric fields, and $\delta b_\|=\delta B_\|/\bar{B}$ is the ratio of the compressive amplitude of the magnetic fluctuations to the mean field. The divergence of this expression is the contribution of the compressive part of the magnetic turbulence. It vanishes for parallel propagation contributing only for propagation $\vec{k}=\vec{k}_\perp$ perpendicular to the mean field. Combining all terms produces the equation
\begin{eqnarray}\label{eq-19}
\frac{\partial}{\partial t}\frac{(\delta b_\|)^2}{2\mu_0}=&-&{\sigma}^T_{\|}(\delta{v}_\perp^\perp)^2\Big[1+(\delta b_\|)^2\Big]\nonumber\\[-2ex] 
&&\\[-2ex]
&-&\frac{1}{\mu_0}\nabla_\perp\cdot\Big[\delta\vec{v}_\perp^\perp\delta b_\|(1+\delta b_\|)\Big]\nonumber
\end{eqnarray}
for the magnetically compressive component. Experimentally it is simple matter to separate out $\delta\vec{B}_\|$. We do not invest further into any discussion of this case. }

\section{Discussion and conclusions}
{Poynting's theorem provides additional information about turbulence which so far had not yet been exploited. It allows to account for the relativistic effect in the electric field and reduces it to a measurement of the turbulent velocity and magnetic fields as suggested by Eq. (\ref{eq-14}). This cannot be circumvented by no means. It is interesting to briefly discuss more recent measurements of electric field, velocity and also density fluctuations  \citep{bale2005,chen2011,podesta2006,podesta2007,podesta2010,podesta2011a,podesta2011b,roberts2010} in this light.}

{The specifications of section 4 show that, as expected from electrodynamics, replacing the electric fluctuations in electromagnetic turbulence the magnetic and velocity fields become related. This follows from relativity. The electric fluctuation field plays an intermediate role of an mediator only. The versions of Poynting's theorem given above explicate the interrelation. They can be applied to stationary homogeneous turbulence providing expressions for the spectrum of the turbulent conductivity as a functional of the magnetic and velocity power spectral densities similar to those given previously {\citep{treumann2017}} but expressed here in terms of the velocity fields.  There we insisted on the independent determination of the electric and magnetic power spectral densities. It turns out that determination of the spectrum of turbulent velocities on all scales is more important.}

{Observations in the solar wind on comparably large scales indicate that the velocity and magnetic spectra in the inertial MHD range exhibit different slopes \citep{podesta2006,podesta2007,podesta2011a}. Velocity power spectra are typically flatter, of slope $-\frac{3}{2}$ (2-D or Kraichnan), than magnetic spectra at 1 AU, which are close to 3-D-Kolmogorov $-\frac{5}{3}$ with apparently less power in the kinetic than in the magnetic energy fluctuations. In fact, there is no obvious reason that they should be similar. Any magnetic fluctuations $\delta\vec{B}$ are, through Amp\`ere's law, related to fluctuations of the electric current 
\begin{equation}
\delta\vec{J}=e\bar{N}(\delta\vec{v}_i-\delta\vec{v}_e)+e\delta N(\bar{\vec{v}}_i-\bar{\vec{v}}_e)
\end{equation}
 assuming quasi-neutrality in turbulence. An example are diamagnetic currents in pressure gradients. Under stationary conditions this reduces to pressure balance. It is the difference in the fluctuations of the ion and electron velocities and the density fluctuations which both contribute. At long MHD scales the average velocities cancel, the last term in the current disappears, but in the first term
%. Hence, since $\vec{k}\cdot\delta\vec{B}=0$ one has dimensionally
%\begin{equation}
%k_\perp^2\big(\delta\vec{B}\big)^2_{\vec{k_\perp}}\sim 2\mu_0\big(\delta\vec{J}\big)^2_{\vec{k_\perp}}=2\mu_0\bar{N}\big(\delta\vec{v}_i-\delta\vec{v}_e\big)^2_{\vec{k_\perp}}
%\end{equation}
the ion and electron velocity fluctuations are not aligned and contribute differently to the spectra. Measured fluctuations in the flow $\delta\vec{v}$ have little in common with the fluctuations of the current. At short scales the second term on the right in the current contributes through the density fluctuations which are caused mainly by fluctuations of the plasma pressure and thus are related to the transverse magnetic pressure. With increasing solar distance in the solar wind the velocity spectra though in the inertial scale range, still being of lower spectral density than the magnetic spectra, seem to approach the Kolmogorov slope \citep{roberts2010} while at the same time intensify. If confirmed, a simple explanation is that in solar wind turbulence the effect of decreasing magnetic field on the flow weakens with increasing solar distance thus gradually loosing dominance.} 

\subsection{Data based thermodynamic considerations}
{The above measurements of the turbulent solar wind velocity spectrum was restricted to the  MHD frequency range $\lesssim 10^{-2}$ Hz. More recent observations \citep{safrankova2013,safrankova2016} based on sophisticated technique aboard the Spektr-R spacecraft, extended to higher frequencies into the range $\lesssim 2$ Hz, presumably scales below the ion gyroradius, where ion kinetic effects become important, for instance in supporting kinetic Alfv\'en waves, and the ions demagnetize.} 

{These measurements confirm the $\sim -\frac{3}{2}$ slope of the turbulent velocity spectrum in the MHD range at frequencies below the ion cyclotron frequency (scales, presumably longer than the ion gyro- and/or inertial scales) thus being more 2-D and flatter than the observed about Kolmogorov-turbulent magnetic spectra. At their higher frequencies they partially cover the kinetic non-magnetized ion range spectra and exhibit power laws of steeper slope close to $-3$ indicating that the turbulent (ion) velocity fluctuations enter a different, presumably still inertial fluid regime when decoupling from the magnetic field. Currents which contribute to the magnetic fluctuations here are carried by magnetized electrons  either perpendicular, as drift currents in the density and temperature gradients of the turbulent eddies thereby forming narrow scale current filaments, or along the magnetic field as kinetic Alfv\'en waves \citep{alexandrova2013}. Signatures of the proximity to this regime are visible as undulations in the velocity spectrum above say $3\times10^{-2}$ Hz already where they form a weak bump on the spectrum which is even more expressed in the density spectrum \citep[first observed already by][their Figure 1]{celnikier1983} which, in general, does not follow neither Kraichnan's nor Kolmogorov's prescription.} 

{It is also of interest that, in the inertial range, the temperature spectrum mimics the velocity spectrum \citep{safrankova2016}. Thus inertial range kinetic energy $\mathrm{d}\epsilon$ and thermal energy $\mathrm{d}T$ follow each other. Assuming ideal gas conditions implies that 
\begin{equation}
\mathrm{d}\epsilon=c_v\mathrm{d}T
\end{equation}
 Therefore, the specific heat $c_v\approx$\,const (within the uncertainty of the measurements) does not change across the inertial range. Such processes are isentropic with 
\begin{equation}
T\sim N^{\gamma-1}
\end{equation}
 where $\gamma=c_p/c_v$ is the ratio of specific heats. Using the average inertial range slopes \citep[see][Fig. 1]{safrankova2016} we then find from the general adiabatic (isentropic) equation \citep[cf., e.g.][]{kittel1980}
\begin{equation}
\mathrm{d}\,\log T-(\gamma -1)\,\mathrm{d}\,\log N=0
\end{equation}
that in the solar wind inertial range {the ratio of specific heats as determined from the fluctuations in density and thermal speed \citep{safrankova2016} is} $\gamma\approx 1.82$, which implies that {under the ideal gas assumption one finds from the relation  
\begin{equation}
\gamma = \frac{2+D}{D}
\end{equation}
between $\gamma$ and the number of dimensions $D$ \citep[cf., e.g.,][]{kittel1980,landau1994} that the inertial range has \emph{fractal dimension} $D\approx 2.46$ which again implies deterministic chaos, self-organization and structure formation \citep[cf., e.g.,][]{barnsley1988,eckmann1985,eckmann1986,zaslavsky1985} } in this range. Since, at least in part of the inertial range, the density and magnetic spectra behave similarly, this reasoning also applies to the turbulent magnetic field.} 

{Entering the ion-kinetic range at higher frequencies, the temperature adjusts to the steeper slope of $\sim -\frac{5}{2}$ suggesting non-adiabaticity and heating over the velocity spectrum as is of course expected when ion-kinetic processes like heating by kinetic Alfv\'en wave turbulence take over in this range.}

\subsection{Application to Alfv\'enic solar wind turbulence}
{It would be desirable to apply the above published measurements to our theoretical determination of the conductivity spectrum. Unfortunately, however, the experimental spectral energy densities are available only in frequency space. Application of the Taylor hypothesis to transfer them into wavenumber space implies imposing a linear Galilean transformation relation $\omega=\bar{\vec{v}}\cdot\vec{k}$ which may hold for very high nonrelativistic average speeds (see also the brief discussion below) and thus in frequency-wavenumber space restricts to multiplication of the conductivity spectrum with a Dirac function $\delta(\omega-\bar{\vec{v}}\cdot\vec{k})$. In the Alfv\'enic  turbulence case one may formally obtain from the measurements of, say, \citet{safrankova2016} and using Eq. (\ref{eq-15a}) that
\begin{equation}
\sigma_{\perp\omega\vec{k}}\sim \Big(1-2\bar{B}\alpha\frac{k_\|}{\omega}\Big)\omega^{-(s_{\delta B}-s_{\delta v}-1)}\delta(\omega-\bar{\vec{v}}\cdot\vec{k})
\end{equation}
where $s_{\delta v},s_{\delta B}$ are the respective experimental slopes of the velocity and magnetic field spectra. Since these are about $\sim\frac{3}{2}$ and $\sim\frac{5}{3}$ respectively, the {conductivity spectrum in the inertial range is also power law of index (up to the factor in brackets and the Dirac function) $s_\sigma=s_{\delta B}-s_{\delta v}-1\approx -\frac{5}{6}$ indicating an \emph{increase} in conductivity $\sigma_{\perp\omega\vec{k}}\sim \omega^{5/6}\delta(\omega-\bar{\vec{v}}\cdot\vec{k})$ with frequency (shrinking temporal scale). Applying the Dirac function which the Taylor hypothesis in addition imposes yields the wavenumber dependence 
\begin{equation}
\sigma_{\perp\vec{k}}\sim \Big(1-\alpha\bar{B}\frac{k_\|}{\bar{\vec{v}}\cdot\vec{k}}\Big)\big(\bar{\vec{v}}\cdot\vec{k}\big)^\frac{5}{6}
\end{equation}
(Note that $k_\|$ refers to the mean magnetic field while in the denominator the wavenumber is parallel to the average flow through Taylor's hypothesis which artificially reintroduces $\bar{\vec{v}}$  at this late place after developing the theory!) If this finding is confirmed and applies, the inertial range turbulent resistance drops with frequency and wave number, meaning that the inertial range in Alfv\'enic turbulence behaves increasingly \emph{less} dissipative towards shorter scales. The system is collisionless, so this \emph{contradicts} the expected self-organisation and structure formation (formation of progressively shorter scale current filaments, eddies etc.) which we have  inferred above from fundamental thermodynamic arguments without making any reference to any additional  hypothesis. This should not be the case. So this result may provide a strong argument against the application of the Taylor hypothesis at least at  short scales, i.e. large wave numbers and frequencies. For the above mentioned reasons concerning observations, such a conclusion must, however, be taken with care.} }

{At this point a  {general} remark on the use of Taylor's hypothesis is in place. It not only reduces the wavenumber-frequency spectrum to the inclusion of a delta function, it also reduces the ``turbulent dispersion relation'' to a linear relation. This might indeed hold as long as the flow velocity is very high, $|\bar{\vec{v}}|\gg \sup{|\delta\vec{v}|}$, a trivial condition. Instead, the ``correct'' turbulent dispersion relation for magnetic turbulence is given through the frequency-wavenumber spectrum of the turbulent conductivity \citep[see][]{treumann2017}. In addition, the Taylor hypothesis applies only to turbulent structures which propagate \emph{along} the mean flow such that $\vec{k}\|\bar{\vec{v}}$. Any turbulence propagating at angle, for instance the rotational velocity component of a turbulent eddy, is thus affected only up to an angle where the projection of the mean speed of the flow onto the wavenumber vector still by far exceeds the turbulent speed. Any strictly perpendicular wave is not affected by Taylor's hypothesis and thus principally cannot become transformed into wavenumber space.} 

{The above used observations make no difference between the propagation directions. Thus any distinction is impossible and any application of spatial scales like gyroradii and inertial scales is questionable because it applies only to part of the mixture of components which makes up the spectra. In order to cure this problem, observations should be split into components perpendicular and parallel to $\bar{\vec{v}}$ and the Taylor hypothesis should be applied to the parallel component only.}

\subsection{Conclusions}
{In the previous section we applied Poynting's theorem to derive expressions between the turbulent conductivity and measurable spectral energy densities. These expressions are formulated in terms of the magnetic and velocity spectra. The electric field appears just on an intermediate step becoming eliminated by the relativistic transformation. These expressions may be useful  in  application to observations but require precise measurements of the velocity field fluctuations. This is the main experimental difficulty. Their knowledge is of general interest in turbulence theory as they allow construction of a turbulent dispersion relation which is not a solution of an eigenmode equation but determines the relation between observed frequencies and wavenumbers. This should provide a useful experimental input into the conventional approach to both fully developed strong \citep{zhou2004,zank2012} and weak \citep{yoon2007a,yoon2007b,boldyrev2009} stationary and homogeneous magnetohydrodynamic turbulence.} 

{Finally we note that we did not use Elsasser \citep{elsasser1950} variables here, the mixed magnetic and flow fields which are usually used in magnetohydrodynamic turbulence theory \citep{biskamp2003,zhou2004,zank2012}. Reformulation of the results in these variables is simple matter. This will be left for a separate investigation.}

\begin{acknowledgement}
This work was part of a Visiting Scientist Programme in 2007 at the International Space Science Institute Bern. We acknowledge interest of the ISSI directorate and the friendly hospitality of the ISSI staff. We thank the ISSI technical administrator Saliba F. Saliba for help, and the librarians Andrea Fischer and Irmela Schweitzer for access to the library and literature. We thank the anonymous reviewer for the constructive critical comments and for directing our attention to some recent publications on measurements of solar wind turbulent velocity and density spectra.
\end{acknowledgement}

\noindent\emph{Data availability.} No data sets were used in this article.

\noindent\emph{Competing interests.} The authors declare that they have no conflict of interest.

\end{document}